\documentclass[aps,twocolumn,amsmath,amssymb,superscriptaddress,preprintnumbers]{revtex4-2}
\usepackage[pdftex]{graphicx}
\pdfoutput=1
\usepackage[outdir=./]{epstopdf}
\usepackage{enumerate,ulem,color}

\newcommand{\sgn}{\mathrm{sgn}}

\newcommand{\beginsection}[1]{\noindent \textit{#1} --- }

\begin{document}
\title{
    Emergence of inductance and capacitance from topological electromagnetism
    }
\author{Yasufumi Araki}
\author{Jun’ichi Ieda}
\affiliation{Advanced Science Research Center, Japan Atomic Energy Agency, Tokai 319-1195, Japan}

\begin{abstract}
Topological electromagnetism owing to nontrivial momentum-space topology of electrons in insulators
gives rise to diverse anomalous magnetoelectric responses.
While conventional inductors and capacitors are based on classical electromagnetism described by Maxwell's equations,
here we show that
topological electromagnetism in combination with spin dynamics in magnets
also generates an inductance or a capacitance.
We build a generic framework to extract the complex impedance on the basis of topological field theory,
and demonstrate the emergence of an inductance or a capacitance in several heterostructure setups.
In comparison with the previously-studied emergent inductances in metallic magnets,
insulators highly suppress
the power loss, because of the absence of Joule heating.
We show that the inductance from topological electromagnetism is achieved at low current and high frequency,
and is also advantageous in its power efficiency,
as characterized by the high quality factor ($Q$-factor).
\end{abstract}

\maketitle


\begin{figure*}[tbp]
    \includegraphics[width=15cm]{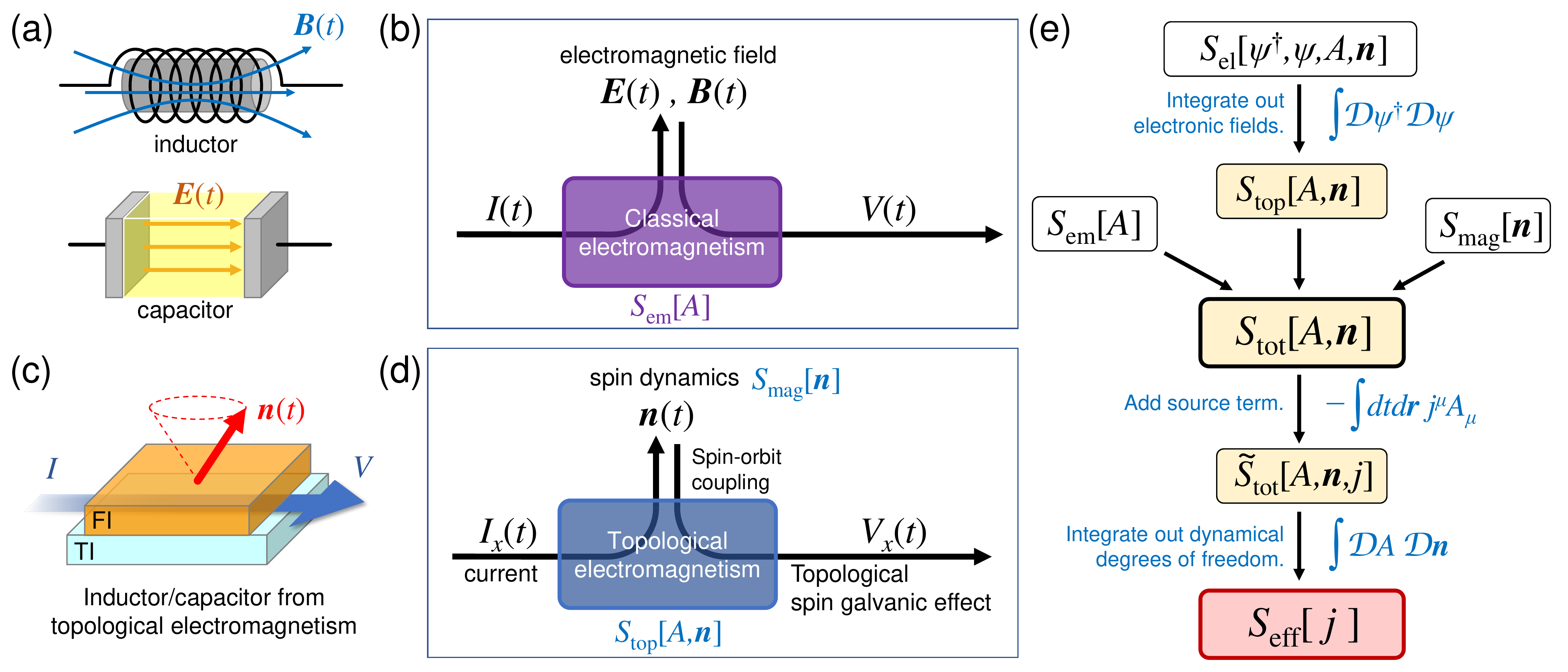}
    \caption{
        \textbf{Schematic images of conventional and topological electromagnetism:}
        (a) Illustrations of conventional inductors and capacitors based on classical electromagnetism.
        The inductance arises from the dynamics of magnetic field $\boldsymbol{B}(t)$,
        and the capacitance arises from the dynamics of electric field $\boldsymbol{E}(t)$.
        (b) Theoretical structure of conventional inductors and capacitors,
        which consists of the Maxwell's classical electromagnetic action $S_{\mathrm{em}}[A]$.
        (c) Illustration of an inductor or a capacitor based on topological electromagnetism.
        As an example, we take a heterostructure of a topological insulator (TI) and a ferromagnetic insulator (FI),
        and make use of the magnetization dynamics $\boldsymbol{n}(t)$.
        See Fig.~\ref{fig:TI-FI}(a) for the detailed setup.
        (d) Theoretical structure of inductors and capacitors based on topological electromagnetism,
        which consists of the topological magnetoelectric action $S_{\mathrm{top}}[A,\boldsymbol{n}]$
        and the action for spin dynamics $S_{\mathrm{mag}}[\boldsymbol{n}]$.
        (e) Image of our field-theoretical framework to derive inductance and capacitance emerging from topological electromagnetism.
        Effective action $S_{\mathrm{eff}}[j]$ 
        describing inductance and capacitance for the charge degrees of freedom $j$
        can be derived from the topological magnetoelectric action $S_{\mathrm{top}}[A,\boldsymbol{n}]$,
        which involves the dynamics of electromagnetic fields $A$ and spins (magnetization) $\boldsymbol{n}$.
    }
    \label{fig:schematic}
\end{figure*}

Topological field theory of electromagnetic fields describes anomalous magnetoelectric responses in materials
that cannot be described by the conventional Maxwell's equations \cite{Qi_2008,Ryu_2012,Niemi_1983,Redlich_1984,Wilczek_1987,Zhang_1989,Essin_2009}.
For instance, the surface state of a three-dimensional (3D) topological insulator (TI) \cite{Fu_2007,Hasan_2010} with magnetism
shows the intrinsic anomalous Hall effect (AHE) \cite{Nomura_2011,Yu_2010,Chang_2013} and the universal magneto-optical effects \cite{Tse_2010,Aguilar_2012,Dziom_2017},
which are described by the $(2+1)$D Chern--Simons theory \cite{Jackiw_1984,Garate_2010,Nomura_2010}.
The edge state of a 2D quantum spin Hall insulator (QSHI) \cite{Murakami_2004,Kane_2005,Bernevig_2006}
is capable of the quantized charge pumping \cite{Qi_2008_2,Chen_2010,Mahfouzi_2010},
which is described by the topological action known as the $\theta$-term in $(1+1)$D \cite{Goldstone_1981}.
Such anomalous magnetoelectric responses are realized by the electronic states with nontrivial band topology in materials,
which we term here as topological electromagnetism.
Notably,
the topological magnetoelectric responses emerging in insulators are free from energy dissipation by Joule heating.
Such nondissipative magnetoelectric responses are
capable of
designing power-saving electronics and spintronics devices,
such as magnetic memories using the topologically induced spin torques
\cite{Garate_2010,Yokoyama_2010,Pesin_2012,Wu_2021,Araki_2021,Yamanouchi_2022}.

In particular, in a circuit component,
the power efficiency
is characterized by the quality factor ($Q$-factor),
\begin{align}
    Q_\omega = \left|\frac{\mathrm{Im} Z_\omega}{\mathrm{Re} Z_\omega}\right|.
\end{align}
Here $Z_\omega$ is the complex impedance of the circuit component,
at the operation frequency $\omega$.
To
achieve a high $Q$-factor,
one needs to reduce the power loss from the resistance $\mathrm{Re} Z_\omega$
and to enhance the reactance $\mathrm{Im} Z_\omega$.
However,
inductors and capacitors,
which are the most fundamental elements broadly used in electric circuits,
are still based on classical electromagnetism [see Figs.~\ref{fig:schematic}(a) and (b)] \cite{Jackson}.
Inductors use the dynamics of
magnetic fields threading magnetic cores in coils,
and capacitors use the dynamics of electric fields in dielectric media
between metallic plates.
Since they operate with conduction currents in metals,
power loss by internal resistance is inevitable,
which reduces the $Q$-factor.

In this work, we establish a theory of inductors and capacitors
based on topological electromagnetism in insulators,
by making use of
topological electronic systems and spin dynamics [see Fig.~\ref{fig:schematic}(c)].
The electrons in topological materials show strong spin-charge coupling
due to the band inversion by spin-orbit coupling (SOC) \cite{Hasan_2010,Kane_2005,Fu_2007,Bernevig_2006}.
As a result of the spin-charge coupling,
the system significantly shows the dynamical current-voltage response,
which can be regarded as an inductance or a capacitance [see Fig.~\ref{fig:schematic}(d)].
Moreover, owing to the dissipationless nature of topological electromagnetism,
power loss is much suppressed and a high $Q$-factor is achieved,
in comparison with conventional inductors and capacitors.

In order to study such
inductive and capacitive behaviors in insulators,
we build a framework
to derive an effective field theory
that describes the current-voltage response of the system,
which is schematically illustrated in Fig.~\ref{fig:schematic}(e).
In this framework, we combine topological electromagnetism
with the dynamics of spins and electromagnetic fields,
and integrate out the microscopic degrees of freedom.
As a result, we reach the effective action
for the collective dynamics of electric current,
which can be directly compared with the classical action for electric circuits \cite{Wells_1938,Agarwal}. 
To demonstrate the emergence of an inductance and a capacitance,
we take a heterostructure composed of a 3D TI and a ferromagnetic insulator (FI) \cite{Wei_2013,Alegria_2014,Jiang_2015,Tang_2017,Yasuda_2017},
and apply our framework.
We chacaterize the inductance appearing here in comparison with the recently reported ``emergent inductance'',
which is the inductive behavior caused by spin dynamics in magnetic metals \cite{Nagaosa_2019,Kurebayashi_2021,Ieda_2021,Yamane_2022,Yokouchi_2020,Kitaori_2021},
by focusing on their operation frequencies and $Q$-factors.
Due to the absence of metallic conduction in TI,
the internal resistance $\mathrm{Re}Z_\omega$ is largely suppressed in the TI-FI heterostructure,
giving rise to a high
$Q_\omega \approx 10 \text{--} 100$.
Moreover, the inductance of the TI-FI heterostructure is available up to the frequency of the ferromagnetic resonance (FMR) in the FI,
which is around $\omega \approx 1 \text{--} 10 \ \mathrm{GHz}$.
This is in a clear contrast with the previously reported emergent inductors using magnetic textures \cite{Nagaosa_2019,Kurebayashi_2021,Ieda_2021,Yokouchi_2020,Kitaori_2021},
whose operation frequencies are limited up to $10 \text{--} 100 \:\mathrm{kHz}$
due to the pinning of magnetic textures.
We conclude that TIs are capable of realizing an inductance with a high $Q$-factor in the high frequency regime.


\section{Results}

\beginsection{Theoretical framework.}
We first show the theoretical framework to treat the contribution of topological electromagnetism to the complex impedance,
the process of which is summarized in Fig.~\ref{fig:schematic}(e).
We formulate the coupled dynamics of
the electric current (or the electromagnetic fields) and the magnetization,
in the Lagrangian formalism of quantum field theory.
We take $\hbar = 1$ throughout this article.
Classical electromagnetism described by the Maxwell's action,
\begin{align}
    S_{\mathrm{em}}[A] &= \frac{1}{2}\int dt d\boldsymbol{r} \left[ \epsilon|\boldsymbol{E}|^2 - \mu^{-1} |\boldsymbol{B}|^2 \right],
\end{align}
accounts for conventional inductance and capacitance
as illustrated in Fig.~\ref{fig:schematic}(a).
Here $A = (A_0,\boldsymbol{A})$ is the gauge potential for
the electric field $\boldsymbol{E} = -\boldsymbol{\nabla} A_0 -\partial_t \boldsymbol{A}$
and the magnetic field $\boldsymbol{B} = \boldsymbol{\nabla}\times \boldsymbol{A}$,
with the dielectric permittivity $\epsilon$ and the magnetic permeability $\mu$.

Topological electromagnetism in materials
comes from the coupling between the electromagnetic fields and topological electron systems.
Moreover, if the system contains a magnetic ordering,
the magnetic moments also participate in topological electromagnetism
via the coupling to the electron spins,
as illustrated in Fig.~\ref{fig:schematic}(d).
Here we denote the effect of topological electromagnetism symbolically as the topological action $S_{\mathrm{top}}[A,\boldsymbol{n}]$,
where the symbol $\boldsymbol{n}$ charactetizes the magnetic ordering.
For the clarity of discussion,
here we take a ferromagnet with the magnetization pointing to the direction $\boldsymbol{n}$.
Microscopically, $S_{\mathrm{top}}$ can be derived from the action $S_{\mathrm{el}}[\psi^\dag,\psi,A,\boldsymbol{n}]$
for the electron system coupled with the electromagnetic fields and the magnetization.
By integrating out the fields $(\psi^\dag,\psi)$ of the electrons,
\begin{align}
    S_{\mathrm{top}}[A,\boldsymbol{n}] &= -i \ln \int \mathcal{D}\psi^\dag \mathcal{D}\psi \ e^{ iS_{\mathrm{el}}[\psi^\dag,\psi,A,\boldsymbol{n}] },
\end{align}
we obtain the topological action $S_{\mathrm{top}}[A,\boldsymbol{n}]$.
The form of $S_{\mathrm{top}}$ depends on the dimensionality and symmetry of the system,
which has been thoroughly studied in the context of topological field theory \cite{Qi_2008,Ryu_2012}.

In addition,
the magnetic ordering $\boldsymbol{n}$ is subject to
the precessional dynamics of spins,
which we symbolically denote by the action $S_{\mathrm{mag}}[\boldsymbol{n}]$.
If the fluctuation around the ground-state magnetic ordering is sufficiently small,
the dynamics can be described
by the bosonic fields of
spin-wave excitations (magnons).

Thus, in topological materials,
the coupled dynamics of the electromagnetic fields $A$ and the magnetic ordering $\boldsymbol{n}$ is described by the action
\begin{align}
    S_{\mathrm{tot}}[A,\boldsymbol{n}] &= S_{\mathrm{em}}[A] + S_{\mathrm{mag}}[\boldsymbol{n}] + S_{\mathrm{top}}[A,\boldsymbol{n}].
\end{align}
From this action $S_{\mathrm{tot}}$,
we now explore the collective dynamics of the electrons,
namely, the dynamics of electric charge and current.
As the field variable conjugate to the electromagnetic field $A$,
we introduce 4-vector field $j = (\rho,\boldsymbol{j})$,
with the charge density $\rho$ and the current density $\boldsymbol{j}$.
By adding the source term,
\begin{align}
    \tilde{S}_{\mathrm{tot}}[A,\boldsymbol{n},j] &= S_{\mathrm{tot}}[A,\boldsymbol{n}] - \int dt d\boldsymbol{r} \ j^\mu A_\mu,
\end{align}
and integrating out the dynamical degrees of freedom $(A,\boldsymbol{n})$,
we reach the effective action $S_{\mathrm{eff}}[j]$ which describes the dynamics of $\boldsymbol{j}$,
\begin{align}
    S_{\mathrm{eff}}[j] &= -i \ln \int \mathcal{D}A \mathcal{D}\boldsymbol{n} \ e^{ i\tilde{S}_{\mathrm{tot}}[A,\boldsymbol{n},j] }.
\end{align}
This formal solution is the main result of this work.
From this form, we can understand the inductive and capacitive behavior of the system.
Up to bilinears in $j$,
it can be compared with the Lagrangian forms of a conventional inductor $(L)$ and capacitor $(C)$ \cite{Wells_1938,Agarwal},
\begin{align}
    S_{L}[I] = \frac{1}{2}L\int dt \ I^2 (t), \quad
    S_{C}[I] = -\frac{1}{2C} \int dt \ P^2(t) ,
\end{align}
where $I(t)$ is the current flowing in an inductor of the inductance $L$,
and $P(t) = \int dt \ I(t)$ is the electric polarization in a capacitor of the capacitance $C$.
By this way,
we obtain the comprehensive expression of the impedance,
including inductance $L$ and capacitance $C$,
which emerge from
the interplay of topological electromagnetism and spin dynamics.

\begin{figure*}[tbp]
    \includegraphics[width=18cm]{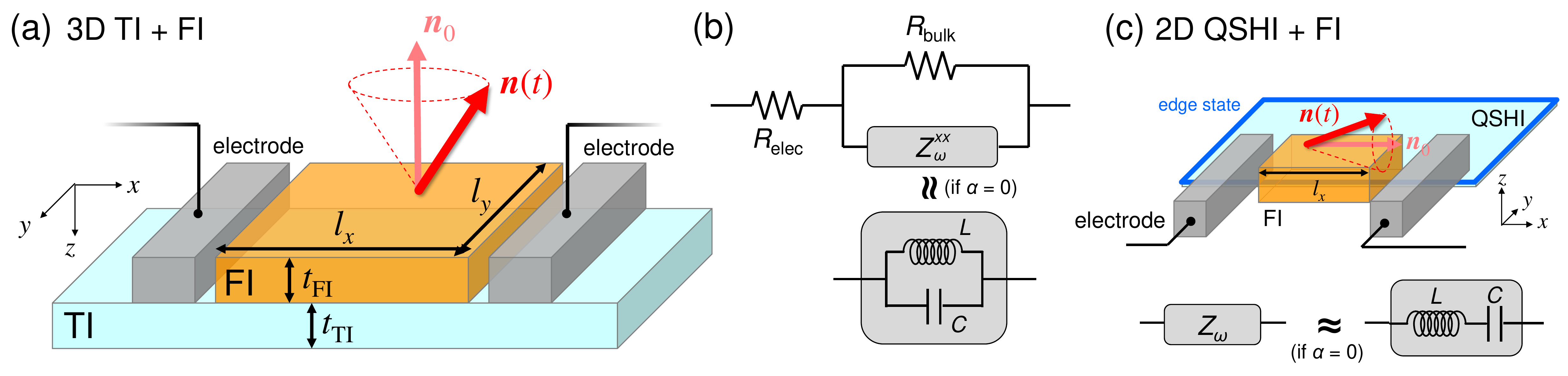}
    \caption{
        \textbf{Schematic illustrations for topological inductors and capacitors:}
        (a) The setup using the two-dimensional (2D) interface of a topological insulator (TI) and a ferromagnetic insulator (FI).
        A current and a voltage are applied longitudinally (in $x$-direction) via the metallic electrodes.
        The magnetization $\boldsymbol{n}(t)$ in the FI precesses around the ground-state position $\boldsymbol{n}_0 = -\hat{\boldsymbol{z}}$.
        The interface is of the length $l_x$ and the width $l_y$,
        and the thicknesses of the TI and FI films are defined as $t_{\mathrm{TI}}$ and $t_{\mathrm{FI}}$, respectively.
        (b) The equivalent circuit for the impedance of the setup (a).
        In addition to the complex impedance $Z^{xx}_\omega$ of the TI-FI interface,
        we take into account the resistance of the bulk $R_{\mathrm{bulk}}$ and that of the electrodes $R_{\mathrm{elec}}$,
        and evaluate the total impedance $Z^{\mathrm{tot}}_\omega$ given by Eq.~(\ref{eq:equivalent-circuit}).
        In the limit $\alpha=0$, $Z^{xx}_\omega$ becomes equivalent to the impedance of a parallel circuit of an inductor $(L)$ and a capacitor $(C)$.
        (c) The setup using the 1D edge of a quantum spin Hall insulator (QSHI) and a FI.
        A current and a voltage are applied longitudinally (in $x$-direction) via the metallic electrodes.
        The magnetization $\boldsymbol{n}(t)$ in the FI precesses around the ground-state position $\boldsymbol{n}_0 = \hat{\boldsymbol{x}}$.
        The interface is of the length $l_x$,
        and the volume of the FI film is defined as $V_{\mathrm{FI}}$.
        In the limit $\alpha=0$, the impedance $Z_\omega$ of the interface becomes equivalent to the impedance of a serial circuit of an inductor $(L)$ and a capacitor $(C)$.
    }
    \label{fig:TI-FI}
\end{figure*}

\beginsection{Interface of topological insulator and ferromagnetic insulator.}
As a demonstration of the effect of topological electromagnetism on the impedance,
we consider a heterostructure of a 3D TI and a FI as shown in Fig.~\ref{fig:TI-FI}(a)
(see Methods for detalis of the calculation process).
The Dirac electrons on the TI surface obtain a mass gap once they are coupled to the out-of-plane magnetization of the FI \cite{Qi_2008,Qi_2006}.
The topological field theory
arising from $(2+1)$D massive Dirac fermions
is the Chern--Simons action \cite{Jackiw_1984,Garate_2010,Nomura_2010},
\begin{align}
    S_{\mathrm{top}}[A,\boldsymbol{n}] &= \frac{\sigma_H}{2} \int dt d^2\boldsymbol{r} \ \epsilon^{\mu\nu\lambda} \mathcal{A}_\mu \partial_\nu \mathcal{A}_\lambda,
    \label{eq:Chern-Simons}
\end{align}
where $\sigma_H = -\frac{e^2}{2\pi} \sgn (n_z)$ is the half-quantized Hall conductivity,
with the elemental charge $e(>0)$,
and $\epsilon^{\mu\nu\lambda}$ with $\mu,\nu,\lambda = 0,x,y$ is the Levi--Civita symbol in $(2+1)$D.
Here $\mathcal{A}$ is the gauge field coupling to the Dirac electrons at the interface.
Due to the spin-momentum locking structure at the TI-FI interface,
the spatial (in-plane) components of $\mathcal{A}$ consist of both the electromagnetic fields $\boldsymbol{A}$ and the magnetization $\boldsymbol{n}$,
as
$\boldsymbol{\mathcal{A}} = \boldsymbol{A} -\tfrac{J}{ev}\hat{\boldsymbol{z}} \times \boldsymbol{n}$.
$J$ parametrizes the interfacial exchange coupling strength betweeen the electron spin and magnetization $\boldsymbol{n}$,
and $v$ denotes the Fermi velocity of the Dirac electrons.
This Chern--Simons action accounts for
the AHE,
yielding the Hall current
$\boldsymbol{j} = \sigma_H \hat{\boldsymbol{z}} \times \boldsymbol{E}$
\cite{Nomura_2011,Yu_2010,Chang_2013}.
Moreover,
there are cross terms of $\boldsymbol{A}$ and $\boldsymbol{n}$ in the action $S_{\mathrm{top}}[A,\boldsymbol{n}]$,
which can be regarded as the effective coupling mediated by the topological electrons.
It is proposed that such an effective coupling
leads to the electric charging of magnetic textures \cite{Nomura_2010},
and also to the electrically induced dynamics of magnetization,
which is known as the topological inverse spin-galvanic effect \cite{Garate_2010,Yokoyama_2010}.

The effective action $S_{\mathrm{eff}}[\boldsymbol{j}]$ for the interfacial current $\boldsymbol{j}$ is derived by the field-theoretical framework
which we have explained above.
For the clarity of the present discussion,
we take the ground-state magnetization direction to the out-of-plane direction $\boldsymbol{n}_0 = -\hat{\boldsymbol{z}}$.
By integrating out the dynamical fields $A$ and $\boldsymbol{n}$, 
we reach the effective action in the bilinear form,
\begin{align}
    S_{\mathrm{eff}}[\boldsymbol{j}] &= l_x l_y \int \frac{d\omega}{2\pi} \frac{z^{\mu\nu}_\omega}{2i\omega} j^\mu_{-\omega} j^\nu_{\omega}, \quad (\mu,\nu = x,y)
    \label{eq:S-eff-j}
\end{align}
where $\boldsymbol{j}_\omega$ denotes the Fourier component of $\boldsymbol{j}(t)$ of the frequency $\omega$,
and $l_{x,y}$ represents the lateral size of the TI-FI interface in each ($x$ or $y$) direction.
The tensor $z_\omega^{\mu\nu}$ relates the electric field and the current density,
$E_{\mu,\omega} = i\omega A_{\mu,\omega} = z_\omega^{\mu\nu} j_\omega^\nu$,
which serves as the complex impedance of a unit area at frequency $\omega$.
Therefore, the impedance of the interface of the area $l_x l_y$ 
is given as $Z_\omega^{\mu\nu} = z_\omega^{\mu\nu} ({l_\mu l_\nu}/{l_x l_y})$
(without summation over $\mu,\nu$).
Note that the system shows both longitudinal and transverse impedances.
In particular, the longitudinal impedance reads
\begin{align}
    Z^{xx}_\omega &= z_0 \frac{i \tilde{\omega}}{(1+i\alpha\tilde{\omega})^2 -\tilde{\omega}^2} \frac{l_x}{l_y t_{\mathrm{FI}}}, \label{eq:TI-impedance}
\end{align}
where $\tilde{\omega} = \omega / \Omega_0$ is the frequency rescaled by the FMR frequency $\Omega_0$,
and $\alpha$ is the parameter that characterizes the relaxation of magnetization dynamics (Gilbert damping) in the FI.
The coefficient $z_0 \equiv (\mu_B / M_s) \left(J/ev\right)^2$,
which has the dimension of volume resistivity,
is governed by the material parameters of the TI and the FI.
$M_s$ is the saturation magnetization of the FI, and $\mu_B$ is the Bohr magneton.
For instance, let us take a heterostructure of $\mathrm{(Bi,Sb)_2 (Te,Se)_3}$ as a TI
and yttrium iron garnet (YIG) as a FI thin films.
By using the parameters $M_s \approx 20 \mu_B \:\mathrm{nm}^{-3}$
for YIG \cite{Anderson_1964,Mallmann_2013},
$J \approx 1 \:\mathrm{eV}$ and $v \approx 10^5 \:\mathrm{m/s}$ for TI-FI interface \cite{Kurebayashi_2014},
we obtain $z_0 \approx 4.5 \times 10^3 \:\mathrm{\mu\Omega}\:\mathrm{cm}$.

\begin{figure}[tbp]
    \includegraphics[width=8cm]{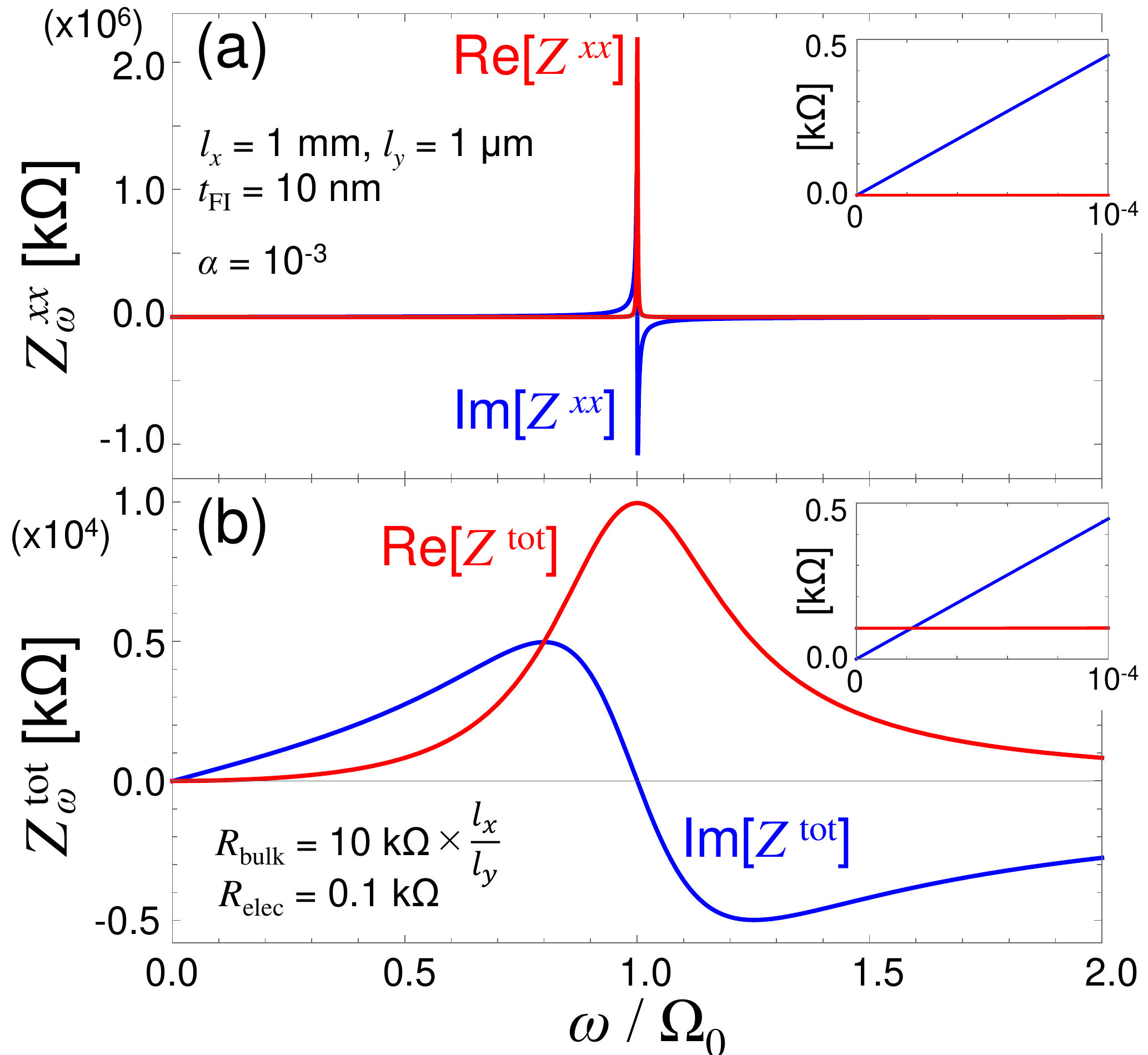}
    \caption{
        \textbf{Frequency dependences of the impedance at TI-FI heterostructure:}
        (a) The longitudinal impedance $Z^{xx}_\omega$ of the interface given by Eq.~(\ref{eq:TI-impedance}).
        We have taken the system size $l_x = 1 \:\mathrm{mm}$, $l_y = 1 \:\mathrm{\mu m}$, $t_{\mathrm{FI}} = 10\:\mathrm{nm}$.
        (b) The impedance of the whole system $Z^{\mathrm{tot}}_\omega$ given by Eq.~(\ref{eq:equivalent-circuit}),
        including the bulk resistance $R_{\mathrm{bulk}} = 10 \:\mathrm{k\Omega} \times (l_x/l_y)$ and $R_{\mathrm{elec}} = 0.1\:\mathrm{k\Omega}$.
        The insets show the magnified views of the impedance at low frequency.
    }
    \label{fig:z-omega}
\end{figure}

The impedance $Z^{xx}_\omega$ shows the resonance structure at $\Omega_0$,
as shown in Fig.~\ref{fig:z-omega}(a).
Such a resonance structure in the electric circuit comes from the magnetic resonance driven by an AC voltage,
which is investigated in Ref.~\cite{Tang_2022}.
In the limit $\alpha = 0$,
the impedance $Z^{xx}_\omega$ becomes compatible to that of an $LC$-parallel circuit shown in Fig.~\ref{fig:TI-FI}(b),
\begin{align}
    Z^{xx}_\omega \approx \left[ \frac{1}{i\omega L} + i\omega C \right]^{-1},
\end{align}
with the inductance and capacitance
\begin{align}
    L = \frac{z_0}{\Omega_0} \frac{l_x}{l_y t_{\mathrm{FI}}}, \quad
    C = \frac{1}{z_0 \Omega_0} \frac{l_y t_{\mathrm{FI}}}{l_x} .
\end{align}
At low frequency up to the FMR frequency $(|\omega| \ll \Omega_0)$,
the impedance shows the inductor-like behavior,
$Z^{xx}_\omega \approx i\omega L$.
The system becomes inductive although the 3D bulk and the 2D interface are insulating,
which can be traced back to the 1D conductive edge channels that contribute to the quantum anomalous Hall state.
$L$ is proportional to the system length $l_x$ and inversely proportional to the cross section $l_y t_{\mathrm{FI}}$,
which is the common behavior seen in the emergent inductors proposed in previous literature \cite{Nagaosa_2019,Kurebayashi_2021,Ieda_2021,Yamane_2022,Yokouchi_2020,Kitaori_2021}.
Thus, they exhibit a large inductance within a small cross section,
in contrast to the conventional inductors formed by coils.
For the system size $l_x = 1 \:\mathrm{mm}, \ l_y = 1 \:\mathrm{\mu m}, \ t_{\mathrm{FI}} = 10\:\mathrm{nm}$,
we can estimate $L = 360 \:\mathrm{\mu H}$.
This $L$ is much larger than the emergent inductances estimated and reported so far using metallic magnets,
which is due to the strong spin-momentum locking at the TI-FI interface.

The real part of $Z^{xx}_\omega$ for the TI-FI interface becomes proportional to the Gilbert damping constant $\alpha$,
$\mathrm{Re} Z^{xx}_\omega \approx \alpha \tilde{\omega}^2 z_0 (l_x/l_y t_{\mathrm{FI}}) $ for $|\omega| \ll \Omega_0$.
It implies that the energy dissipation occurs exclusively from the relaxation of magnetization dynamics in the FI.
Due to the small damping constant in the FI
($\alpha \approx 10^{-4}\text{--}10^{-3}$),
we expect that the power loss in the present system is ideally well suppressed
in comparison with the conventional and emergent inductors using metals.
The sharp resonance structure seen in Fig.~\ref{fig:z-omega}(a) accounts for the suppression of power loss.

If the measurement is done at room temperature,
the internal resistance $R_{\mathrm{bulk}}$ from the bulk conduction becomes nonnegligible \cite{Chang_2013,Wang_2017,Wu_2021,Fujimura_2021},
which works in parallel to the impedance $Z^{xx}_\omega$ from the interface electrons.
Moreover,
the resistance $R_{\mathrm{elec}}$ of the metallic electrodes and leads is also present in the experimental measurement,
which is serially connected to $Z^{xx}_\omega$ and $R_{\mathrm{bulk}}$.
Therefore, we now consider an equivalent circuit shown in Fig.~\ref{fig:TI-FI}(b).
To quantify the power efficiency in the whole system,
we evaluate its impedance,
\begin{align}
    Z^{\mathrm{tot}}_\omega &= R_{\mathrm{elec}} + \frac{1}{\left(Z^{xx}_\omega\right)^{-1} + \left(R_{\mathrm{bulk}}\right)^{-1}},
    \label{eq:equivalent-circuit}
\end{align}
with the system size defined above.
We have taken
the approximate values of the bulk resistivity
$R_{\mathrm{bulk}} = 10 \ \mathrm{k\Omega} \times (l_x/l_y)$ seen at room temperature in Bi-based TIs \cite{Chang_2013,Wang_2017,Wu_2021,Fujimura_2021},
the Gilbert damping constant $\alpha = 10^{-3}$ seen in YIG \cite{Haidar_2016},
and fixed
$R_{\mathrm{elec}} = 0.1 \:\mathrm{k\Omega}$ in Fig.~\ref{fig:z-omega}(b).
In the low-frequency regime $(|\omega| \ll \Omega_0)$,
the reactance $\mathrm{Im} Z^{\mathrm{tot}}_\omega$ rises linearly in $\omega$ like an inductor,
whereas the resistance $\mathrm{Re} Z^{\mathrm{tot}}_\omega$ is dominated by the electrodes $R_{\mathrm{elec}}$.
On the other hand, around the FMR frequency $\omega \approx \Omega_0$,
the impedance is dominated by the bulk resistance $R_{\mathrm{bulk}}$.
Here the reactance component $\mathrm{Im} Z^{\mathrm{tot}}_\omega$ gets suppressed and the resonance becomes smeared.
Therefore, the TI-FI heterostructure exhibits the inductive behavior up to the FMR frequency $\omega \lesssim \Omega_0$.

\beginsection{1D edge of quantum spin Hall insulator.}
As another example,
let us breifly mention the case of a 2D 
QSHI.
A QSHI has spin-helical edge states, which are described as 1D Dirac electrons \cite{Murakami_2004,Kane_2005,Bernevig_2006}.
We consider the case where these edge states within the length $l_x$
are coupled to an FI,
as shown in Fig.~\ref{fig:TI-FI}(c).
We take the volume of the FI $V_{\mathrm{FI}}$,
and its magnetization direction $\boldsymbol{n}$ fluctuating around the ground-state magnetization $\boldsymbol{n}_0 = \hat{\boldsymbol{x}}$.
The Dirac electrons on the edge become gapped and contributes to the $(1+1)$D topological action
\begin{align}
    S_{\mathrm{top}}[A_x,\boldsymbol{n}] &= -e \int dt dx \frac{\theta}{2\pi} \partial_t \mathcal{A}_x,
    \label{eq:theta-1D}
\end{align}
known as the $\theta$-term \cite{Qi_2008_2,Goldstone_1981},
with
$\theta = \arctan (n_y/n_x) \approx n_y$
if the fluctuation of $\boldsymbol{n}$ around $x$-axis is small.
In a manner similar to the $(2+1)$D case,
the gauge field $\mathcal{A}_x = A_x - \tfrac{J}{ev} n_z$ contains both the electromagnetic field and the magnetization,
with $J$ the exchange coupling strength and $v$ the velocity of the Dirac electrons on the 1D edge.
Based on this topological action,
we can evaluate the impedance of this QSHI-FI system (see Supplementary Information for the calculation process).
In the limit
$\alpha = 0$,
the effective impedance $Z_\omega$ becomes compatible to that of an $LC$-serial circuit as shown in Fig.~\ref{fig:TI-FI}(c),
\begin{align}
    Z_\omega &\approx i\omega L + \frac{1}{i\omega C},
\end{align}
with
\begin{align}
    C = \frac{e^2}{ 4\pi^2 N_s} \frac{1}{\Omega_0}, \quad 
    L = \frac{ 4\pi^2 N_s}{e^2} \frac{r^2}{\Omega_0}.
    \label{eq:QSHI-impedance}
\end{align}
Here $N_s = V_{\mathrm{FI}} M_s /\mu_B$ is the number of magnetic moments in the FI,
and $r = 1+ \frac{J l_x}{2\pi v N_s}$ is the renormalization factor for the spin Berry phase
which arises from the topological magnetoelectric coupling in Eq.~(\ref{eq:theta-1D}).
In contrast to the $(2+1)$D case, this system behaves like a capacitor at low frequency below the FMR frequency $\Omega_0$,
and hence the system becomes totally insulating in the limit $\omega=0$.
This is reasonable,
because the present QSHI-FI heterostructure does not have any conducting edge channel,
and the electric current is pumped only by the magnetization dynamics driven at finite frequency.

\section{Discussion}

\begin{figure}[tbp]
    \includegraphics[width=8cm]{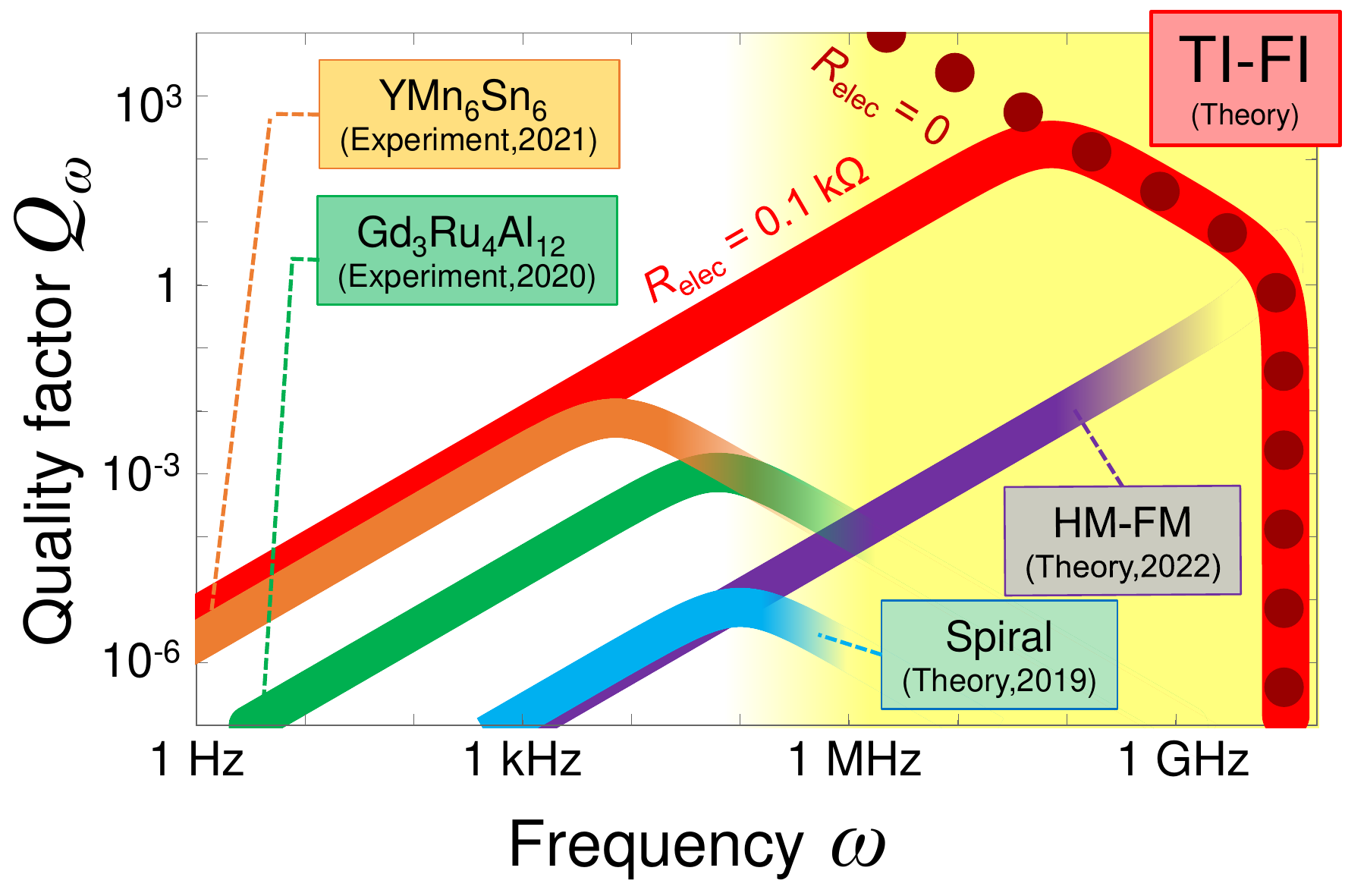}
    \caption{
        \textbf{Comparison of the performances of emergent inductors:}
        The operation frequencies $\omega$ and the $Q$-factors $Q_\omega$ as functions of $\omega$ are plotted
        for various emergent inductor setups using magnetization dynamics.
        The red branches (TI-FI) account for the heterostructure of a topological insulator and a ferromagnetic insulator investigated in this work.
        The red solid line is our estimation with the internal resistance $R_{\mathrm{elec}} = 0.1 \:\mathrm{k\Omega}$ of the electrodes,
        while the dark-red dotted line is the estimation without $R_{\mathrm{elec}}$.
        The other lines show the theoretically estimated and experimentally measured data of the emergent inductors seen in the previous works (see Discussion section and Supplementary Information for the details and references for the previous works).
        The TI-FI heterostructure is advantageous in realizing a high $Q$-factor in the yellow-shaded frequency region $(\omega \gtrsim 100 \:\mathrm{kHz})$.
    }
    \label{fig:q-factors}
\end{figure}

Let us now focus on the inductive behavior obtained above.
We compare its operation frequency $(\omega)$ and $Q$-factor $(Q_\omega)$,
with those of the emergent inductors using metallic magnets reported in previous literature,
which are summarized in Fig.~\ref{fig:q-factors}.
(See Supplementary Information for details of the displayed data.)


In metallic magnets with spin textures,
such as magnetic spirals,
the interplay of the electric current and the dynamics of the spin textures,
known as the spin-transfer torque \cite{Slonczewski_1989,Berger_1996,Tserkovnyak_2005} and the spinmotive force \cite{Berger_1986,Volovik_1987,Barnes_2007},
gives rise to the emergent inductance \cite{Nagaosa_2019,Kurebayashi_2021,Ieda_2021}.
The emergent inductance was measured in the helimagnetic states of $\mathrm{Gd_3 Ru_4 Al_{12}}$ \cite{Yokouchi_2020} and $\mathrm{Y Mn_6 Sn_6}$ \cite{Kitaori_2021},
$L \approx 100 \:\mathrm{nH} \text{--} 1 \:\mathrm{\mu H}$ at frequencies $1 \text{--} 10 \:\mathrm{kHz}$.
They show large inductances and high $Q$-factors due to the short helical pitches of a few nanometers.
However,
the inductances are suppressed at frequencies $\omega \gtrsim 10\:\mathrm{kHz}$.
Theoretically,
the suppression of emergent inductance is associated to the weakening of extrinsic pinning at high frequency.

In a heterostructure of heavy metal (HM) and ferromagnetic metal (FM),
interplay of the current and the magnetization dynamics is governed by SOC,
and the emergent inductance arises even if the magnetization is uniform \cite{Yamane_2022}.
The operation frequency is not limited by the depinning effect,
and it reaches as high as the FMR frequency.
In the heterostructure of Co and Pt,
the emergent inductance of $L \approx 0.1 \:\mathrm{nH}$ is estimated for the area of $1 \:\mathrm{\mu m} \times 1\:\mathrm{\mu m}$ and the thickness of $10 \mathrm{nm}$,
up to the order of $\mathrm{GHz}$.
On the other hand, due to the metallicity in the bulk,
the inductive behavior accompanies
a large conduction current that is unnecessary for the inductive behavior.
Such a large operation current
leads to the Joule heating by the bulk resistivity,
which reduces the $Q$-factor below unity.

In comparison with the emergent inductors listed above,
the topological electromagnetism in TIs is advantageous
in realizing an emergent inductor with a high operation frequency,
a low opeation current, and a high $Q$-factor.
Since the emergent inductors with TIs use dynamics of uniform magnetization,
the operation frequency can reach the FMR frequency of the order of $1 \text{--} 10 \ \mathrm{GHz}$,
which is much higher than those of the spiral magnets.
Compared with the HM-FM heterostructures,
the bulk conduction that is unnecessary for the emergent inductance is much suppressed in TIs,
and hence the emergent inductors using TIs can operate with a lower current and a higher $Q$-factor.
These advantages
can be understood within the unified framework shown here based on the topological field theory.

While topological field theory has been intensely studied from its geometrical aspects,
its practical advantage in describing physical phenomena is not well explored so far.
Our theoretical
study clearly demonstrates its new practical advantage,
in understanding and designing 
power-efficient inductors and capacitors using topological materials.
While we have incorporated the dynamics of ferromagnetic ordering $\boldsymbol{n}(t)$ in Fig.~\ref{fig:schematic}(e) to demonstrate our field-theoretical treatment of the inductance and capacitance,
$\boldsymbol{n}(t)$ can be replaced with any other types of orderings
that couple to electron systems.
For instance, recent theoretical and experimental studies have discovered a class of antiferromagnetic insulators,
called the axion insulators,
whose magnetoelectric responses are described by the topological action known as the $\theta$-term in $(3+1)$D \cite{Sekine_2021,Wilczek_1987,Li_2010,Mong_2010}.
The dynamics of antiferromagnetic ordering (N\'{e}el vector) may give rise to topological inductance and capacitance therein.
Moreover, our theoretical framework may not be limited to magnetic orderings,
but can also be extended to charge orderings and superconductivity,
which is left for future studies.

\section{Methods}
\beginsection{Calculation process at TI-FI interface.}
The electrons at the TI surface are described by the 2D Dirac Hamiltonian with spin-momentum locking,
$\mathcal{H}_{\mathrm{el}}^0 = v (\boldsymbol{p} \times \boldsymbol{\sigma})_z$,
where $v$ denotes the Fermi velocity at low energy,
$\boldsymbol{p}=-i\boldsymbol{\nabla}$ is the momentum operator,
and $\boldsymbol{\sigma}$ is the Pauli matrix representing the electron spin.
Here we assume that the bulk of the TI and FI is fully insulating
and will not consider the bulk conduction.
The surface Dirac electrons are
coupled to the electromagnetic fields $A$ by the minimal coupling $-j^\mu A_\mu$,
and to the magnetization $\boldsymbol{n}$ of the FI by 
the exchange coupling $J \boldsymbol{n} \cdot \boldsymbol{\sigma}$ (with the coupling constant $J>0$).
Then the action $S_{\mathrm{el}}[\psi^\dag,\psi,A,\boldsymbol{n}]$ reads
\begin{align}
    S_{\mathrm{el}}[\psi^\dag,\psi,A,\boldsymbol{n}] =
    \int dt d^2\boldsymbol{r} \ \psi^\dag \left[ \pi_0 - v(\boldsymbol{\pi} \times \boldsymbol{\sigma})_z - J n_z \sigma_z \right] \psi,
\end{align}
where the momentum operators are defined by
\begin{align}
    \pi_0 \equiv i\partial_t - e A_0, \quad
    \boldsymbol{\pi} \equiv \boldsymbol{p} - e\boldsymbol{\mathcal{A}} \equiv -i\boldsymbol{\nabla} - e\left[\boldsymbol{A} -\tfrac{J}{ev}\hat{\boldsymbol{z}} \times \boldsymbol{n}\right]. \nonumber
\end{align}
Note that the in-plane components of the magnetization $\boldsymbol{n}$ leads to the shift of the momentum and acts like the vector potential.
On the other hand,
the out-of-plane component $n_z$ opens a gap at the Dirac point
and leads to the massive Dirac spectrum $E_\pm(\boldsymbol{k}) = \pm \sqrt{(vk)^2 + (Jn_z)^2}$.
This mass gap gives rise to the momentum-space Berry curvature $\Omega^\pm_z(\boldsymbol{k}) = \mp\frac{J n_z}{2|E_\pm(\boldsymbol{k})|^3}$ for the eigenstates $E_\pm(\boldsymbol{k})$,
which is the source of the topological magnetoelectric responses.
If the Fermi level lies inside the gap of the massive Dirac spectrum,
the whole system, including both the bulk and the interface, becomes insulating.
By integrating out the fermionic fields,
the topological action $S_{\mathrm{top}}[A,\boldsymbol{n}]$ results in the $(2+1)$D Chern--Simons form shown by Eq.~(\ref{eq:Chern-Simons}).

In addition to $S_{\mathrm{top}}$,
we formulate the dynamics of the magnetization $S_{\mathrm{mag}}$ and the electromagnetic fields $S_{\mathrm{em}}$.
The fluctuation of magnetization around the ground-state position,
$\boldsymbol{u} = \boldsymbol{n} - \boldsymbol{n}_0$
(with $\boldsymbol{n}_0 = -\hat{\boldsymbol{z}}$),
is described as magnons.
Here we assume that the magnetization is formed by the spins with the magnitude $S_0$ for each spin,
distributed at the number density $\nu_s$.
Then the saturation magnetization $M_s$ becomes $M_s = \nu_s \mu_B S_0$.
The spin variable $\boldsymbol{S}(\boldsymbol{r},t)$ is related to $\boldsymbol{n}(\boldsymbol{r},t)$ as $\boldsymbol{S} = -S_0 \boldsymbol{n}$.
By the Holstein--Primakoff transformation,
the fluctuation of spin $\boldsymbol{S}(\boldsymbol{r},t)$ around the ground-state position $S_0 \hat{\boldsymbol{z}}$ is described by the magnonic field $\phi(\boldsymbol{r},t)$,
with the relation
\begin{align}
    S_x + i S_y &=  \phi \left[2S_0 - \phi^* \phi\right]^{1/2}.
\end{align}
Therefore, $\phi$ is related to $\boldsymbol{u}$ as
\begin{align}
    \phi = -\sqrt{\frac{S_0}{2}} \left[u_x+iu_y + O(S_0^{-1})\right].
\end{align}
In terms of $\phi$,
the magnetization dynamics is described by the effective action
\begin{align}
    S_{\mathrm{mag}}[\boldsymbol{n}] &= \nu_s \int_{\mathrm{FI}} dt d^3\boldsymbol{r} \ \phi^* [i\partial_t - \Omega(\boldsymbol{p})] \phi,
\end{align}
where $\Omega(\boldsymbol{p})$ denotes the magnon dispersion as a function of momentum $\boldsymbol{p}$.
The integral runs over the volume of the FI.
Note that the relaxation of magnetization dynamics is omitted in this action,
which shall be restored in the discussion later on.

For the electromagnetic fields,
we focus on the dynamics induced by the AC current,
and hence we omit the magnetic components and keep the electric components,
\begin{align}
    S_{\mathrm{em}}[\boldsymbol{A}] &= \frac{\bar{\epsilon}}{2} \int_{\mathrm{full}} dt d^3\boldsymbol{r} \ |\boldsymbol{E}|^2,
\end{align}
where we take the Coulomb gauge $\boldsymbol{E}(\boldsymbol{r},t) = -\partial_t \boldsymbol{A}$.
The spatial integral runs over the whole system including both the TI and the FI,
with $\bar{\epsilon}$ the dielectric permittivity averaged over the whole system.

From $S_{\mathrm{tot}} = S_{\mathrm{top}} + S_{\mathrm{mag}} + S_{\mathrm{em}}$ defined above,
we now evaluate the effective action $S_{\mathrm{eff}}$ for the electric current.
Here we take the thin-film geometry for the FI and TI,
with their thicknesses $t_{\mathrm{FI}}$ and $t_{\mathrm{TI}}$, respectively,
and bring them into contact at the interface with the length $l_x$ and the width $l_y$,
as shown in Fig.~\ref{fig:TI-FI}(a). 
Over the whole system, we assume the spatial homogeneity of the fields $\boldsymbol{A}$ and $\boldsymbol{n}(=-\hat{\boldsymbol{z}} +\boldsymbol{u})$.
By this assumption, the total action $S_{\mathrm{tot}}$ can be rearranged in the matrix form for $\boldsymbol{A}$ and $\boldsymbol{u}$,
\begin{align}
    S_{\mathrm{tot}}[\boldsymbol{A},\boldsymbol{u}] &= l_x l_y \int \frac{d\omega}{2\pi}  
    \begin{pmatrix}
        A_\omega^* & u_\omega^*
    \end{pmatrix}
    \hat{K}_\omega
    \begin{pmatrix}
        A_\omega \\ u_\omega
    \end{pmatrix},
\end{align}
with the Fourier components $A_\omega \equiv A_{x,\omega} + i A_{y,\omega}$ and $u_\omega \equiv u_{x,\omega} + i u_{y,\omega}$
labeled by the frequency $\omega$.
The components of the kernel matrix $\hat{K}_\omega$ read
\begin{align}
    & \hat{K}_\omega =
    \begin{pmatrix}
        K_\omega^{AA} & K_\omega^{Au} \\
        K_\omega^{uA} & K_\omega^{uu}
    \end{pmatrix}
    ,
    \\
    & K_\omega^{AA} = \frac{1}{2}\bar{\epsilon} t_{\mathrm{full}}\omega^2 + \frac{\sigma_H}{2} \omega, \quad 
    K_\omega^{Au} = \left( K_\omega^{u A}\right)^* = i\lambda \frac{\sigma_H}{2} \omega, \nonumber \\
    & K_\omega^{uu} = \tfrac{1}{2} t_{\mathrm{FI}} \nu_s S_0 (\omega - \Omega_0 - i\alpha|\omega|) + \lambda^2 \frac{\sigma_H}{2} \omega,
\end{align}
with $t_{\mathrm{full}} = t_{\mathrm{TI}} + t_{\mathrm{FI}}$.
Here $\Omega_0 \equiv \Omega(\boldsymbol{p} =0)$ is the FMR frequency,
and $\lambda \equiv J/ev$ is the material parameter characterizing the effect of exchange coupling at the interface.
We have restored the effect of the Gilbert damping $\alpha$,
by shifting $\Omega_0 \rightarrow \Omega_0 + i\alpha|\omega|$.
At the frequency $\omega$ lower than the electrostatic energy scale $\Omega_{\mathrm{el}} = {e^2}/{4\pi \bar{\epsilon} t_{\mathrm{full}}}$,
the dielectric part $\frac{1}{2}\bar{\epsilon} t_{\mathrm{full}}\omega^2$ in $K_\omega^{AA}$ becomes negligibly small compared to the topological part $(\sigma_H/2)\omega$.
Since $\Omega_{\mathrm{el}}$ is at the order of $1\:\mathrm{THz}$ in the films of $t_{\mathrm{full}} \approx 10\:\mathrm{nm}$,
which is much higher than the FMR frequency,
we neglect the dielectric part in the following discussion.

By adding the source term to this action,
we have
\begin{align}
    & \tilde{S}_{\mathrm{tot}}[\boldsymbol{A},\boldsymbol{u},\boldsymbol{j}] = S_{\mathrm{tot}}[\boldsymbol{A},\boldsymbol{u}] -l_x l_y \int \frac{d\omega}{2\pi} \boldsymbol{j}_{-\omega} \cdot \boldsymbol{A}_{\omega} \\
    &=
    l_x l_y \int \frac{d\omega}{2\pi}  
    \begin{pmatrix}
        A_\omega^* & u_\omega^* & j_\omega^*
    \end{pmatrix}
    \left(
        \begin{array}{cc|c}
            K_\omega^{AA} & K_\omega^{Au} & -\frac{1}{2} \\
            K_\omega^{uA} & K_\omega^{uu} & 0 \\ \hline
            -\frac{1}{2} & 0 & 0
        \end{array}
    \right)
    \begin{pmatrix}
        A_\omega \\ u_\omega \\ j_\omega
    \end{pmatrix},
    \nonumber
\end{align}
with $j_\omega = j_\omega^x + i j_\omega^y$.
By integrating out the dynamical degrees of freedom $A_\omega$ and $\phi_\omega$,
we obtain the effective action for the 2D current density $\boldsymbol{j}$ in the bilinear form,
\begin{align}
    S_{\mathrm{eff}}[\boldsymbol{j}] &= l_x l_y \int \frac{d\omega}{2\pi} \Gamma_\omega j_\omega^* j_\omega,
\end{align}
with
\begin{align}
    \Gamma_\omega = -\frac{1}{4} \left[\hat{K}_\omega^{-1}\right]^{AA}
    = -\frac{1}{4} \frac{K_\omega^{uu}}{K_\omega^{AA} K_\omega^{uu} - K_\omega^{Au} K_\omega^{uA}}.
\end{align}
By decomposing the real and imaginary parts of $j_\omega$,
we obtain the form shown by Eq.~(\ref{eq:S-eff-j}).
The longitudinal and transverse parts of the tensor $z_\omega^{\mu\nu} \ (\omega \geq 0)$ read
\begin{align}
    z_\omega^{xx} = z_\omega^{yy} &= i\omega \left(\Gamma_\omega + \Gamma_{-\omega}\right) \label{eq:z-xx} \\
    &= \frac{\lambda^2}{t_{\mathrm{FI}}\nu_s S_0} \frac{i\omega(\Omega_0+i\alpha \omega)}{(\Omega_0+i\alpha \omega)^2-\omega^2} , \nonumber\\
    z_\omega^{xy} = -z_\omega^{yx} &= -\omega \left(\Gamma_\omega - \Gamma_{-\omega}\right) \\
    &= -\frac{1}{\sigma_H} - \frac{\lambda^2}{t_{\mathrm{FI}}\nu_s S_0} \frac{\omega^2}{(\Omega_0+i\alpha \omega)^2-\omega^2}, \nonumber
\end{align}
respectively.
The transverse part $z_\omega^{xy}$ contains the real constant term $-1/\sigma_H$,
which comes from the quantized AHE intrinsic to the TI surface.
Furthermore, both the longitudinal and transverse parts contain the terms proportional to $\lambda^2 = (J/ev)^2$,
which can be regarded as the impedances emerging from the coupling of the TI and the FI.
From the above form of $z_\omega^{xx}$ given by Eq.~(\ref{eq:z-xx}),
we reach $Z^{xx}_\omega$ shown in Eq.~(\ref{eq:TI-impedance}).

\section*{Acknowledgments}
The authors thank Shunsuke Fukami, Kentaro Nomura, Eiji Saitoh, Kei Yamamoto, and Yuta Yamane for fruitful discussions.
This work is partially supported by KAKENHI (No. 19H05622, No. 20H01830, and No. 22K03538).
Y.A. is supported by the Leading Initiative for Excellent Young Researchers (LEADER).

\section*{Authors contributions}
All authors contributed equally to the results presented in this work and the writing of
the manuscript.


\end{document}


\title{Supplementary Information for \\
``Emergence of inductance and capacitance from topological electromagnetism''}
\author{Yasufumi Araki}
\author{Jun’ichi Ieda}
\affiliation{Advanced Science Research Center, Japan Atomic Energy Agency, Tokai 319-1195, Japan}

\begin{abstract}
\end{abstract}

\maketitle

\section{Derivation of impedance on the edge of quantum spin Hall insulator}
Here we show the detailed calculation process of the topological capacitance and inductance
on the 1D edge of QSHI.
The electrons on the edge of QSHI are described as 1D Dirac fermions.
By incorporating the coupling to the electromagnetic field $A= (A_0,A_x)$ 
and the magnetization $\boldsymbol{n}$ in the FI,
the action $S_{\mathrm{el}}[\psi^\dag,\psi,A_x,\boldsymbol{n}]$ reads
\begin{align}
    S_{\mathrm{el}}[\psi^\dag,\psi,A,\boldsymbol{n}] &= \int dt dx \ \psi^\dag \left[ \pi_0 -v\pi_x \sigma_z - J\boldsymbol{n}_\perp\cdot\sigma_\perp \right] \psi,
\end{align}
where we take the edge along the $x$-axis,
and the quantization axis of spin to the $z$-axis.
The momentum operators are defined as
\begin{align}
    \pi_0 \equiv i\partial_t -eA_0, \quad
    \pi_x \equiv p_x -e\mathcal{A}_x = -i\partial_x -e\left(A_x -\tfrac{J n_z}{ev} \right),
\end{align}
where $v$ denotes the Fermi velocity of the edge electrons.
The in-plane components of the magnetization $\boldsymbol{n}_\perp = n_x \hat{\boldsymbol{x}} + n_y \hat{\boldsymbol{y}}$
opens a gap in the Dirac spectrum of the edge electrons,
and the whole system becomes insulating.
By integrating out the fermionic fields $(\psi^\dag,\psi)$,
we obtain the topological action in $(1+1)$D,
\begin{align}
    S_{\mathrm{top}}[A,\boldsymbol{n}] &= -\int dt dx \frac{\theta}{2\pi} e \mathcal{E}_x,
\end{align}
with $\theta = \arctan(n_y/n_x)$ and $\mathcal{E}_x = \partial_t \mathcal{A}_x - \partial_x A_0$.
This form of $S_{\mathrm{top}}$ is known as the $\theta$-term in the context of topological field theory.
Here we take the ground-state magnetization $\boldsymbol{n}_0 = \hat{\boldsymbol{x}}$
and treat the fluctuations $\boldsymbol{u} = \boldsymbol{n} - \hat{\boldsymbol{x}}$ perturbatively,
which yields $\theta \approx u_y$.
The topological action thus becomes
\begin{align}
    S_{\mathrm{top}}[A,\boldsymbol{u}] &\approx -\frac{e}{2\pi} \int dt dx \ u_y \partial_t\left(A_x-\tfrac{J u_z}{ev}\right), \label{eq:top-1d}
\end{align}
which gives the effective coupling between the electromagnetic fields and the magnetic fluctuations.

In a manner similar to the case of TI-FI interface,
we also need to incorporate the effect of magnetization dynamics $S_{\mathrm{mag}}$
and the classical electromagnetic response $S_{\mathrm{em}}$.
Here we assume that the magnetization in FI is formed by the spins with the magnitude $S_0$ for each spin,
distributed at the number density $\nu_s$.
Then $S_{\mathrm{mag}}$ is given in terms of the magnonic field $\phi$,
\begin{align}
    S_{\mathrm{mag}}[\phi] &= \nu_s \int_{\mathrm{FI}} dt d^3\boldsymbol{r} \ \phi^* \left[ i\partial_t -\Omega_\alpha \right] \phi,
\end{align}
where the spatial integral is taken over the volume $V_{\mathrm{FI}}$ of the FI.
We have assumed the homogeneity of the magnonic field,
and have extracted the Kittel mode $\Omega_0$ for simplicity of discussion.
We have shifted $\Omega_0 \rightarrow \Omega_\alpha \equiv \Omega_0 + i\alpha|\omega|$,
with the Gilbert damping constant $\alpha$.
The magnonic field $\phi(\boldsymbol{r},t)$ is related to the magnetic fluctuations $u_{y,z}(\boldsymbol{r},t)$ by the Holstein--Primakoff transformation,
\begin{align}
    u_y &= -\frac{1}{\sqrt{2S_0}} \left[\phi + \phi^* + O(S_0^{-1})\right], \\ 
    u_z &= -\frac{i}{\sqrt{2S_0}} \left[\phi - \phi^* + O(S_0^{-1})\right], \nonumber
\end{align}
or
\begin{align}
    \phi = \sqrt{\frac{S_0}{2}}\left[ -u_y + i u_z+ O(S_0^{-1})\right] . \quad
\end{align}
Therefore, $S_{\mathrm{mag}}$ reads
\begin{align}
    S_{\mathrm{mag}}[\boldsymbol{u}] &= \frac{\nu_s S_0}{2} \int_{\mathrm{FI}} dt d^3\boldsymbol{r} \left[ u_y \dot{u}_z - u_z \dot{u}_y - \Omega_\alpha (u_y^2 + u_z^2) \right].
\end{align}
On the other hand, $S_{\mathrm{em}}$ is given in terms of the electric field $E_x = \partial_t A_x$,
\begin{align}
    S_{\mathrm{em}}[A_x] &= \frac{1}{2} \bar{\epsilon} \int_{\mathrm{full}} dt d^3\boldsymbol{r} \  E_x^2,
\end{align}
where the spatial integral is taken over the whole insulating system of the volume $V_{\mathrm{full}}$
and $\bar{\epsilon}$ is the averaged dielectric permittivity.

By summing up the three terms given above,
the total action reads
\begin{align}
    S_{\mathrm{tot}}[A_x,\boldsymbol{u}] &= S_{\mathrm{top}}[A_x,\boldsymbol{u}] + S_{\mathrm{mag}}[\boldsymbol{u}] + S_{\mathrm{em}}[A_x] \\
    &= \int_0^{2\pi} \frac{d\omega}{2\pi} \left( A_{x,-\omega} \ u_{y,-\omega} \ u_{z,-\omega} \right) \hat{K}_\omega
    \begin{pmatrix}
        A_{x,\omega} \\ u_{y,\omega} \\ u_{z,\omega}
    \end{pmatrix}
    .
    \nonumber
\end{align}
The $3\times 3$ kernel matrix $\hat{K}_\omega$ is defined as
\begin{align}
    \hat{K}_\omega &=
    \begin{pmatrix}
        K^{A_x A_x}_\omega & K^{A_x u_y}_\omega & K^{A_x u_z}_\omega \\
        K^{u_y A_x}_\omega & K^{u_y u_y}_\omega & K^{u_y u_z}_\omega \\
        K^{u_z A_x}_\omega & K^{u_z u_y}_\omega & K^{u_z u_z}_\omega
    \end{pmatrix}, \\
    K^{A_x A_x}_\omega &= \tfrac{1}{2}\bar{\epsilon} V_{\mathrm{full}}\omega^2 \nonumber\\
    K^{A_x u_y}_\omega = -K^{u_y A_x}_\omega &= -\tfrac{e}{4\pi} l_x i\omega \nonumber\\
    K^{A_x u_z}_\omega = -K^{u_z A_x}_\omega &= 0 \nonumber\\
    K^{u_y u_y}_\omega = K^{u_z u_z}_\omega &= -\tfrac{1}{2} V_{\mathrm{FI}} \nu_s S_0 \Omega_\alpha \nonumber\\
    K^{u_y u_z}_\omega = -K^{u_z u_y}_\omega &= -\tfrac{1}{2} V_{\mathrm{FI}} \nu_s S_0 i\omega -\tfrac{J}{4\pi v} l_x i\omega \nonumber\\
    & \equiv -\tfrac{1}{2} V_{\mathrm{FI}} \nu_s S_0 r i\omega , \nonumber
\end{align}
where $l_x$ is the length of the contact between the QSHI edge and the FI.
The factor $r = 1+\frac{J l_x}{2\pi v V_{\mathrm{FI}}\nu_s S_0}$ leads to the renormalization of the spin Berry phase,
which comes from the topological magnetoelectric coupling in Eq.~(\ref{eq:top-1d}).
By adding the source term of the current $I^x$, the total action becomes
\begin{align}
    \tilde{S}_{\mathrm{tot}}[A_x,\boldsymbol{u},I^x] &= S_{\mathrm{tot}}[A_x,\boldsymbol{u}] -\int dt dx \ I^x A_x \\
    &= \int_0^{2\pi} \frac{d\omega}{2\pi} \left( A_{x,-\omega} \ u_{y,-\omega} \ u_{z,-\omega} \ I^x_{-\omega} \right) \nonumber \\
    & \quad \quad \quad \quad \times \left(
        \begin{array}{ccc|c}
             & & & -\frac{l_x}{2} \\
             & \hat{K}_\omega & & 0 \\
             & & & 0 \\ \hline
             -\frac{l_x}{2} & 0 & 0 & 0
        \end{array}
    \right)
    \begin{pmatrix}
        A_{x,\omega} \\ u_{y,\omega} \\ u_{z,\omega} \\ I^x_{\omega}
    \end{pmatrix}
    .
    \nonumber
\end{align}
By integrating out the dynamical fields $A_x$ and $\boldsymbol{u}$ by the Gaussian integral,
we obtain
\begin{align}
    S_{\mathrm{eff}}[I^x] & = \int_0^{2\pi} \frac{d\omega}{2\pi} \ I^x_{-\omega} \frac{Z_\omega}{2i\omega} I^x_{\omega} \\
    &= -\frac{l_x^2}{4} \int_0^{2\pi} \frac{d\omega}{2\pi} \ I^x_{-\omega} \left[ \hat{K}_\omega^{-1} \right]^{A_x A_x} I^x_{\omega}, \nonumber
\end{align}
where
\begin{align}
    \left[ \hat{K}_\omega^{-1} \right]^{A_x A_x} &= \frac{K^{u_y u_y}_\omega K^{u_z u_z}_\omega - K^{u_y u_z}_\omega K^{u_z u_y}_\omega}{\det \hat{K}_\omega} .
\end{align}
In a manner similar to the case of TI-FI interface,
the dielectric part $K^{A_x A_x}_\omega = \frac{\bar{\epsilon}}{2} V_{\mathrm{full}} \omega^2$ is negligibly small as long as we consider the frequency $\omega$ around the FMR frequency,
and hence we have
\begin{align}
    \left[ \hat{K}_\omega^{-1} \right]^{A_x A_x} &\approx \frac{K^{u_y u_y}_\omega K^{u_z u_z}_\omega - K^{u_y u_z}_\omega K^{u_z u_y}_\omega}{-K^{u_z u_z}_\omega K^{u_y A_x}_\omega K^{A_x u_y}_\omega} \\
    &= \frac{\left( \tfrac{1}{2} V_{\mathrm{FI}} \nu_s S_0 \right)^2 \left( \Omega_\alpha^2 - r^2\omega^2 \right)}{\tfrac{1}{2} V_{\mathrm{FI}} \nu_s S_0 \Omega_\alpha \left(\tfrac{e}{4\pi} l_x \omega\right)^2 } \nonumber \\
    &= \frac{\tfrac{1}{2} V_{\mathrm{FI}} \nu_s S_0}{\left(\tfrac{e}{4\pi} l_x\right)^2} \left[ \frac{\Omega_\alpha}{\omega^2} -\frac{r^2}{\Omega_\alpha} \right]. \nonumber
\end{align}
Therefore, the effective impedance $Z_\omega$ becomes
\begin{align}
    Z_\omega &= 2i\omega \frac{-l_x^2}{4} \left[ \hat{K}_\omega^{-1} \right]^{A_x A_x} \\
    &\approx \frac{ 4\pi^2 V_{\mathrm{FI}} \nu_s S_0}{e^2} \left[ \frac{r^2}{\Omega_\alpha}i\omega +  \frac{\Omega_\alpha}{i\omega} \right]. \nonumber
\end{align}
In the limit $\alpha = 0$,
this form is compatible with the impedance of an $LC$-serial circuit shown in the main text.

\section{Comparison among the emergent inductors}
In this section of Supplementary Information,
we show details of our comparison of the operation frequencies and the $Q$-factors among the emergent inductors using magnetization dynamics,
which is summarized in Fig.~4 in the main text.
We compare the following five cases.

\subsection{TI-FI heterostructure (Theory)}
As we have shown in the main text,
the inductive behavior in the TI-FI heterostructure is available at frequencies lower than the FMR,
$|\omega| \ll \Omega_0$.
In this inductive regime,
we estimate the $Q$-factor of the whole heterostructure system.
We suppose $\mathrm{(Bi,Sb)_2 (Te,Se)_3}$ (BSTS) for the TI layer,
with the Fermi level properly tuned inside the bulk bandgap,
and yttrium iron garnet (YIG) as the FI layer,
from which we obtain $z_0 \approx 4.5 \times 10^3 \:\mathrm{\mu\Omega}\:\mathrm{cm}$ as shown in the main text.

To estimate the impedance of the whole system,
we take the size of the bilayer $l_x = 1\:\mathrm{mm}$, $l_y = 1 \:\mathrm{\mu m}$, and $t_{\mathrm{FI}} = 10 \:\mathrm{nm}$.
The longitudinal sheet resistivity of the bulk of BSTS-YIG heterostructures was measured to be around $\approx 10 \:\mathrm{k\Omega}$,
and hence we take the bulk resistance $R_{\mathrm{bulk}} \approx 10^4 \:\mathrm{k\Omega}$ in our system geometry,
which acts in parallel with the interface impedance $Z^{xx}_\omega$.
As the internal resistance of the electrodes,
we take $R_{\mathrm{elec}} = 0.1\:\mathrm{k\Omega}$,
and then we obtain the behavior of $Z^{\mathrm{tot}}_\omega$ 
as shown in Fig.~3(b) in the main text.

In the low-frequency regime $|\omega| \ll \Omega_0$,
as shown in the inset of Fig.~3(b) in the main text,
the reactance shows the inductive behavior $\mathrm{Im} Z^{\mathrm{tot}}_\omega \propto \omega$,
and the resistance $\mathrm{Re}Z^{\mathrm{tot}}_\omega$ is dominated by the constant value $R_{\mathrm{elec}}$.
Therefore, the $Q$-factor becomes proportional to $\omega$ in the low-frequency regime,
as shown by the red line in Fig.~4 in the main text.

If the internal resistance of the electrodes is omitted $(R_{\mathrm{elec}} =0)$,
the resistance comes from the magnetic relaxation and behaves as $\mathrm{Re} Z^{xx}_\omega \propto \omega^2$ in the low-frequency regime.
Therefore, the $Q$-factor becomes inversely proportional to $\omega$,
as shown by the dark-red dotted line in Fig.~4 in the main text.

\subsection{Spiral magnet (Theory)}
In emergent inductors using metallic magnets,
the conduction current participates in both the impedance $Z^{\mathrm{EI}}_\omega$ from the spinmotive force (SMF)
and the internal resistance $R$ from the electron-phonon or disorder scatterings,
and hence the system can be regarded as a serial circuit of $Z^{\mathrm{EI}}_\omega$ and $R$,
\begin{align}
    Z^{\mathrm{tot}}_\omega = Z^{\mathrm{EI}}_\omega + R.
\end{align}
We estimate the $Q$-factor from this $Z^{\mathrm{tot}}_\omega$ in the following discussions.

In a metallic magnet hosting magnetic textures,
the dynamics of the electric current and the magnetic textures are coupled
by the spin-transfer torque (STT) and the SMF.
When a current is applied to the system,
the dynamics of magnetic texture is driven by the STT.
The magnetic texture eventually returns to the ground-state position due to the pinning,
and its dynamics exerts the SMF on the electrons.
The overall sequence of those processes can be regarded as an inductance.
If the extrinsic pinning by impurities is weaker than the intrinsic pinning by the magnetic anisotropy,
the impedance from the SMF obeys the Debye-type relaxation,
\begin{align}
    Z^{\mathrm{EI}}_\omega = \frac{i\omega L_0}{1- i \frac{\omega}{\omega_D}}.
\end{align}
The upper limit $L_0$ for the emergent inductance of the spiral magnet becomes
\begin{align}
    L_0 = \frac{\pi \gamma l_x}{2 e \lambda A j_{\mathrm{int}}},
\end{align}
as proposed by Ref.~39 in the main text.
Here $\gamma$ is the gyromagnetic ratio,
$\lambda$ is the pitch of the magnetic spiral,
$j_\mathrm{\mathrm{int}}$ is the threshold current density for the intrinsic pinning,
$l_x$ is the length of the sample,
and $A$ is the cross section of the sample.
The Debye frequency $\omega_D$ is given as
\begin{align}
    \omega_D &= \frac{a^3 p}{2e S_0}\frac{\beta j_{\mathrm{imp}}}{\alpha \lambda},
\end{align}
where $a$ is the lattice spacing,
$p$ is the electron spin polarization ratio,
$S_0$ is the magnitude of each spin,
$\alpha$ is the Gilbert damping constant,
$\beta$ is the dimensionless coefficient for the nonadiabatic STT,
and $j_{\mathrm{imp}}$ is the threshold current density for the extrinsic pinning by impurities.

By taking the material parameters
$\lambda = 3\ \mathrm{nm}$,
$a = 1 \ \mathrm{nm}$,
$p = 0.1$,
$S_0 = 1$,
$j_{\mathrm{int}} = 10^{11} \:\mathrm{A} \mathrm{m}^{-2}$,
$j_{\mathrm{imp}} = 10^6 \:\mathrm{A} \mathrm{m}^{-2}$,
and $\alpha \approx \beta$,
the Debye frequency $\omega_D$ is estimated to be $\approx 100 \ \mathrm{kHz}$.
For the sample of the size $l_x = 1\:\mathrm{mm}$ and $A = (1\:\mathrm{\mu m})^2$,
the inductance reaches $L_0 \approx 1.5 \:\mathrm{nH}$.
By taking the resistivity $\rho \approx 1 \:\mathrm{\mu\Omega} \mathrm{cm}$,
which gives $R = 10 \:\mathrm{\Omega}$ for the given $l_x$ and $A$,
the $Q$-factor reaches up to $\approx 10^{-5}$ for $\omega \lesssim \omega_D \approx 100 \:\mathrm{kHz}$,
as shown in Fig.~4 in the main text.

\subsection{HM-FM heterostructure (Theory)}
In the heterostructure of HM and FM,
spin-charge conversion occurs at the interface due to spin-orbit coupling (SOC).
An injected current drives the dynamics of magnetization by the spin-orbit torque (SOT).
The magnetic anisotropy serves as a restoring force on the magnetization,
and the magnetization dynamics exerts the SMF from SOC on the elctrons.
The overall sequence of those processes can be regarded as an inductance.
The impedance $Z^{\mathrm{EI}}_\omega$ from the SMF at the interface shows the resonance structure,
\begin{align}
    Z^{\mathrm{EI}}_\omega = i\omega L_0 \frac{1 + i\alpha\tilde{\omega}}{(1+i\alpha\tilde{\omega})^2 -\tilde{\omega}^2},
\end{align}
as proposed by Ref.~42 in the main text,
where $\tilde{\omega} \equiv \omega / \omega_K$ is the frequency rescaled by the anisotropy frequency $\omega_K$.
The anisotropy frequency is given as $\omega_K = (\gamma/M_s) 2K$,
where $\gamma$ is the gyromagnetic ratio,
$M_s$ is the saturation magnetization,
and $K$ is the energy of easy-axis magnetic anisotropy.
The inductance $L_0$ is given as
\begin{align}
    L_0 = \left(\frac{p m}{e}\right)^2 \frac{l_x}{A}{g^2}{2K},
\end{align}
where $p$ is the electron spin polarization ratio,
$m$ is the effective mass of electron,
$g$ is the coefficient for the Rashba or Dresselhaus SOC,
$l_x$ is the length of the sample,
and $A$ is the cross section of the sample.

By taking the material parameters
$p = 0.5$,
$m = 9.1 \times 10^{-31} \:\mathrm{kg}$,
$g = 10^{-10} \:\mathrm{eV} \mathrm{m}$,
and $\omega_K = 10 \:\mathrm{GHz}$,
the inductance was estimated to be
\begin{align}
    L_0 \approx 1 \ \mathrm{nH} \times \frac{l_x/\mathrm{nm}}{A/\mathrm{nm}^2}.
\end{align}
By taking Pt as the HM layer,
which has the resistivity $\rho \approx 1 \:\mathrm{\mu\Omega} \mathrm{cm}$,
the $Q$-factor reaches up to $\approx 1$ for $\omega \lesssim \omega_K = 10 \:\mathrm{GHz}$,
as shown in Fig.~4 in the main text.

\subsection{$\mathrm{Gd_3 Ru_4 Al_{12}}$ (Experiment)}
The emergent inductance of $\mathrm{Gd_3 Ru_4 Al_{12}}$ was experimentally reported by Ref.~43 in the main text.
In the sample of the length $l_x = 9\:\mathrm{\mu m}$, the width $l_y = 10 \:\mathrm{\mu m}$, and the thickness $t = 300 \:\mathrm{nm}$ (Device 2 shown by Ref.~43 in the main text),
the real part of the inductance was measured to reach $\mathrm{Re} L \approx - 40\ \:\mathrm{nH}$
in its proper-screw state at $T=14.7\:\mathrm{K}$.
The measured inductance shows the Debye relaxation structure,
above the Debye frequency $f_D= \omega_D/2\pi \approx 10 \:\mathrm{kHz}$.
By using the measured resistivity $\rho_{xx} = 40 \:\mathrm{\mu \Omega} \mathrm{cm}$,
the $Q$-factor reaches $Q_\omega \approx 3 \times 10^{-4}$ at $\omega = \omega_D$.

\subsection{$\mathrm{Y Mn_6 Sn_6}$ (Experiment)}
The emergent inductance of $\mathrm{Y Mn_6 Sn_6}$ was experimentally reported by Ref.~44 in the main text.
In the sample of the length $l_x = 28.8\:\mathrm{\mu m}$, the width $l_y = 5.5 \:\mathrm{\mu m}$, and the thickness $t = 1.8 \:\mathrm{\mu m}$ (Device 2 shown by Ref.~44 in the main text),
the real part of the inductance was measured to reach $\mathrm{Re} L \approx - 6\:\mathrm{\mu H}$
in its proper-screw state at $T=100 \:\mathrm{K}$.
The measured inductance shows the Debye relaxation structure,
above the Debye frequency $f_D = \omega_D/2\pi \approx 1 \:\mathrm{kHz}$.
By using the measured resistivity $\rho_{xx} = 70 \:\mathrm{\mu \Omega} \mathrm{cm}$,
the $Q$-factor reaches $Q_\omega \approx 3 \times 10^{-3}$ at $\omega = \omega_D$.
